\documentclass[10pt,twocolumn,letterpaper]{article}

\usepackage{cvpr}
\usepackage{times}
\usepackage{epsfig}
\usepackage{graphicx}
\usepackage{amsmath}
\usepackage{amssymb}

\usepackage{amsmath}
\usepackage{amssymb}
\usepackage{array,multirow}
\usepackage{floatrow}
\usepackage[breaklinks=true,bookmarks=false,pagebackref=true,breaklinks=true,letterpaper=true,colorlinks,bookmarks=false]{hyperref}
\usepackage{float}
\hypersetup{
    colorlinks=true,
    linkcolor=blue,
    filecolor=magenta,      
    urlcolor=cyan,
    pdftitle={Overleaf Example},
    pdfpagemode=FullScreen,
    }

\usepackage[pagebackref=true,breaklinks=true,letterpaper=true,colorlinks,bookmarks=false]{hyperref}

\cvprfinalcopy 

\def\cvprPaperID{****} 

\ifcvprfinal\fi
\begin{document}

\title{Learning to Automatically Diagnose Multiple Diseases in Pediatric Chest Radiographs Using Deep Convolutional Neural Networks}

\author{Thanh T. Tran$^{1}$, Hieu H. Pham$^{1,2, \dag}$, Thang V. Nguyen$^{1}$, Tung T. Le$^{1}$, Hieu T. Nguyen$^{1}$, Ha Q. Nguyen$^{1,2}$ \\
$^{1}$Medical Imaging Center, Vingroup Big Data Institute, Hanoi, Vietnam\\
$^{2}$College of Engineering \& Computer Science, VinUniversity, Hanoi, Vietnam\\
{$\dag$Corresponding author \tt v.hieuph4@vinbigdata.org}
}

\maketitle

\begin{abstract}
   Chest radiograph (CXR) interpretation is critical for the diagnosis of various thoracic diseases in pediatric patients. This task, however, is error-prone and requires a high level of understanding of radiologic expertise. Recently, deep convolutional neural networks (D-CNNs) have shown remarkable performance in interpreting CXR in adults. However, there is a lack of evidence indicating that D-CNNs can recognize accurately multiple lung pathologies from pediatric CXR scans. In particular, the development of diagnostic models for the detection of pediatric chest diseases faces significant challenges such as (\textbf{i}) lack of physician-annotated datasets and (\textbf{ii}) class imbalance problems. In this paper, we retrospectively collect a large dataset of 5,017 pediatric CXR scans, for which each is manually labeled by an experienced radiologist for the presence of 10 common pathologies. A D-CNN model is then trained on 3,550 annotated scans to classify multiple pediatric lung pathologies automatically. To address the high-class imbalance issue, we propose to modify and apply ``Distribution-Balanced loss'' for training D-CNNs which reshapes the standard Binary-Cross Entropy loss (BCE) to efficiently learn harder samples by down-weighting the loss assigned to the majority classes. On an independent test set of 777 studies, the proposed approach yields an area under the receiver operating characteristic (AUC) of 0.709 (95\% CI, 0.690–0.729). The sensitivity, specificity, and \textit{F1}-score at the cutoff value are 0.722 (0.694–0.750), 0.579 (0.563–0.595), and 0.389 (0.373–0.405), respectively. These results significantly outperform previous state-of-the-art methods on most of the target diseases. Moreover, our ablation studies validate the effectiveness of the proposed loss function compared to other standard losses, e.g., BCE and Focal Loss, for this learning task. Overall, we demonstrate the potential of D-CNNs in interpreting pediatric CXRs.
\end{abstract}

\section{Introduction}

Common respiratory pathologies such as pneumonia, chronic obstructive pulmonary disease (COPD), bronchiolitis, asthma, and lung cancer are the primary cause of mortality among children worldwide~\cite{zar2014global}. Each year, acute lower respiratory tract infections (\textit{e.g.}, pneumonia, lung abscess, or bronchitis) cause several hundred thousand deaths among children under five years old~\cite{GBD2015, unicef2006}. Chest radiograph (CXR) is currently the most common diagnostic imaging tool for diagnosing frequent thorax diseases in children. Interpreting CXR scans, however, requires an in-depth knowledge of radiological signs of different lung conditions, making this process challenging, time-consuming, and prone to error. For instance, Swingler \textit{et al}.~\cite{swingler2005diagnostic} reported that the diagnostic accuracy of experienced specialist pediatricians and primary level practitioners in detecting radiographic lymphadenopathy was low, with a sensitivity of 67\% and a specificity of 59\%. Beyond that, the average inter-observer agreement and intra-observer agreement in the CXR interpretation in children were only 33\% and 55\%, respectively~\cite{du2002observer}. Thus, it is crucial to develop computer-aided diagnosis (CAD) systems that can automatically detect common thorax diseases in children and add clinical value, like notifying clinicians about abnormal cases for further interpretation.

Deep learning (DL) has recently succeeded in many biomedical applications, especially detecting chest abnormalities in adult patients~\cite{rajpurkar2017chexnet,rajpurkar2018deep,irvin2019chexpert,baltruschat2019comparison}. Nonetheless, few studies have demonstrated the ability of DL models in identifying common lung diseases in pediatric patients. To the best of our knowledge, most DL-based pediatric CXR interpretation models have focused on a single disease such as pneumonia~\cite{gu2018classification,rajaraman2019visualizing,liang2020transfer} or pneumothorax~\cite{taylor2018automated}. Except the work of Chen \textit{et al}.~\cite{Chen2020}, no work has been published to date on the automatic multi-label classification of pediatric CXR scans. Several obstacles that prevent the progress of using DL for the pediatric CXR interpretation have been reported in Moore \textit{et al}.~\cite{moore2019machine}, in which key challenges for pediatric imaging DL-based computer-aided diagnosis (CAD) development include: (\textbf{1}) acquire pediatric-specific big data sets sufficient for algorithm development; (\textbf{2}) accurately label large volumes of pediatric CXR images; and (\textbf{3}) require the explainable ability of diagnostic models. Additionally, learning with real-world pediatric CXR imaging data also faces the imbalance between the positive and negative samples, making the models more sensitive to the majority classes. To address these challenges, we develop and validate in this study a DL-based CAD system that can accurately detect multiple pediatric lung pathologies from CXR images. A large pediatric CXR dataset is collected and manually annotated by expert radiologists. To address the high-class imbalance issue, we train DL networks with a modified version of “Distribution-Balanced loss” that down-weights the loss assigned to the majority of classes. Our experimental results validate the effectiveness of the proposed loss function compared to other standard losses, and in the meantime, significantly outperform previous state-of-the-art methods for the pediatric CXR interpretation. To summarize, the main contributions of this work are the following:

$\bullet$ We develop and evaluate state-of-the-art D-CNNs for multi-label diseases classification from pediatric CXR scans. To the best of our knowledge, the proposed approach is the first to investigate the learning capacity of D-CNNs on pediatric CXR scans to diagnose 10 types of common chest pathologies.

$\bullet$ We propose modifying and applying the recently introduced Distribution-Balanced loss to reduce the impact of imbalance data issues. This loss function is designed to encourage classifiers to learn better for minority classes and lightens the dominance of negative samples. Our ablation studies on the real-world imbalanced pediatric CXR dataset validated the effectiveness of the proposed loss function compared to the other standard losses.

$\bullet$ The proposed approach surpasses previous state-of-the-art results. The codes and dataset used in this study will be shared as a part of a bigger project that we will release on our project website at \url{https://vindr.ai/datasets/pediatric-cxr}.

\section{Related Works}
\subsection{DL-based for pediatric CXR interpretation}

Several DL-based approaches for pediatric CXR interpretation have been introduced in recent years. However, most of these studies focus on detecting one specific type of lung pathology like pneumonia~\cite{gu2018classification,Rajaraman2019,Labhane2020,Sharma2020,Singh2020}. Most recently, Chen \textit{et al}.~\cite{Chen2020} proposed a DL-based CAD scheme for 4 common pulmonary diseases of children, including \textit{bronchitis}, \textit{bronchopneumonia}, \textit{lobar pneumonia}, and \textit{pneumothorax}. However, this approach was trained and tested on a quite small dataset (\textit{N = 2668}). We recognize that the lack of large-scale pediatric CXR datasets with high-quality images and human experts’ annotations is the main obstacle of the field. To fill this lack, we constructed a benchmark dataset of 5,017 pediatric CXR images in Digital Imaging and Communications in Medicine (DICOM) format. Each image was manually annotated by an experienced radiologist for the presence of 10 types of pathologies. To our knowledge, this is currently the largest pediatric CXR dataset for multi-disease classification task.

\subsection{Multi-label learning and imbalance data issue}
Predicting thoracic diseases from pediatric CXR scans is considered as a multi-label classification problem, in which each input example can be associated with possibly more than one disease label. Many works have studied the problem of multi-label learning, and extensive overviews can be found in Zhang \textit{et al}.~\cite{zhang2013review}, Ganda~\textit{et al}.~\cite{ganda2018survey}, and Liu~\textit{et al}.~\cite{liu2020emerging}. A common approach to the multi-label classification problem is to train a D-CNN model with the BCE loss~\cite{zhang2013review,tsoumakas2007multi}, in which positive and negative classes are treated equally. Multi-label classification tasks in medical imaging are often challenging due to the dominance of negative examples. To handle this challenge, several approaches proposed to train D-CNNs using weighted BCE losses~\cite{ibrahim2020confidence,rajpurkar2017chexnet} instead of the ordinary BCE. In this work, we propose a new loss function based on the idea of Distribution-Balanced loss~\cite{wu2020distribution} to the multi-label classification of pediatric CXR scans. The proposed loss function is based on two key ideas: (\textbf{1}) rebalance the weights that consider the impact caused by label co-occurrence, in particular in the case of absence of all pathologies; and (\textbf{2}) mitigate the over-suppression of negative labels. Our experiments show that the proposed loss achieves remarkable improvement compared to other standard losses (\textit{i.e.}, BCE, weighted BCE, Focal loss, and the original Distribution-Balanced loss) in classifying pediatric CXR diseases.

\section{Methodology}
This section introduces details of the proposed approach. We first give an overview of our DL framework for the pediatric CXR interpretation (Section~\ref{sect:3.1}). We then provide a formulation of the multi-label classification (Section~\ref{sect:3.2}). Next, a new modified distribution-balanced loss that deals with the imbalanced classes in pediatric CXR dataset is described (Section~\ref{sect:3.3}). This section also introduces network architecture choices and training methodology (Section~\ref{sect:3.4} \& Section \ref{sect:3.5}). Finally, we visually investigate model behavior in its prediction of the pathology  (Section~\ref{sect:3.6}).

\subsection{Overall framework \label{sect:3.1}}

\begin{figure}[ht]
\begin{center}
   \includegraphics[width=1\linewidth]{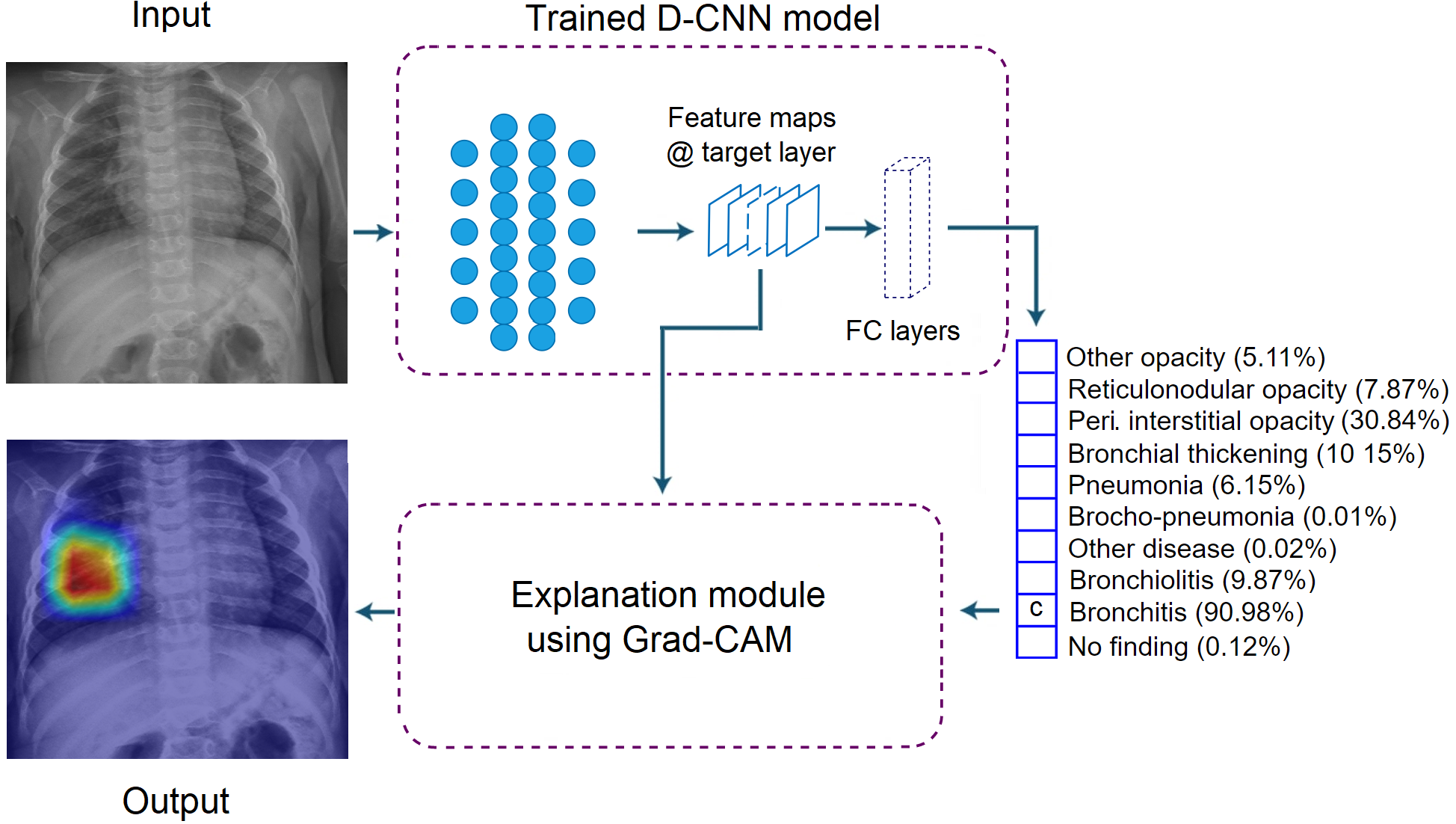}
\end{center}
   \caption{Illustration of our multi-label classification task, which aims to build a DL system for predicting the probability of the presence of 10 different pathologies in pediatric CXRs. The system takes a pediatric CXR as input and outputs the probability of multiple pathologies. It also localizes areas in the image most indicative of the pathology via a heat map created by Grad-CAM method~\cite{Selvaraju2019_gradcam}.}\label{fig:overall}
\end{figure}

The proposed approach is a supervised multi-label classification framework using D-CNNs. It accepts a CXR of children patients as input and predicts the presence of 10 common thoracic diseases: \textit{Reticulonodular opacity, Peribronchovascular interstitial opacity (PIO), Other opacity, Bronchial thickening, Bronchitis, Brocho-pneumonia, Bronchiolitis, Pneumonia, Other disease,} and \textit{No finding}. To train the D-CNNs, a large-scale and annotated pediatric CXR dataset of 5,017 scans has been constructed (Section~\ref{sect:dataset}). With the nature of imbalance among disease labels, the dataset could introduce a bias in favor of the majority diseases. This leads to skew the model performance dramatically. To addresses this challenge, a new loss function that down-weights the loss assigned to majority classes is proposed to train the networks. Finally, a visual explanation module based on  Grad-CAMs~\cite{Selvaraju2019_gradcam} is also used to improve the model’s transparency by indicating areas in the image that are most indicative of the pathology. An overview of the proposed approach is illustrated in Figure~\ref{fig:overall}.

\subsection{Problem formulation \label{sect:3.2}} 
In a multi-label classification setting, we are given a training set $\mathcal{D}$ consisting of \textit{N} samples, $\mathcal{D} = \left\{\left(\textbf{x}^{(i)}, y^{(i)}\right); i = 1, \ldots, N\right\}$ where each input image $\textbf{x}^{(i)} \in \mathcal{X}$ is associated with a multi-label vector $y^{(i)} \in [0, 1]^{\mathcal{C}}$. Here, $\mathcal{C}$ denotes the number of classes. Our task is to learn  a discriminant function $f_{\boldsymbol{\theta}} : \mathcal{X} \to  \mathbb{R}^{\mathcal{C}}$ to make accurate diagnoses of common thoracic diseases from unseen pediatric CXRs. In general, this learning task could be performed by training a D-CNN, parameterized by weights $\boldsymbol{\theta}$ that the BCE loss function is minimized over the training set $\mathcal{D}$. For multi-label classification problem, the sigmoid activation function
$(1 + e^{-z_k})^{-1}$
is applied to the logits $z_k$ at the last layer of the network. The total BCE loss $\mathcal{L}(\boldsymbol{\theta})$ is simple average of all BCE terms over all training examples and given by
\begin{align}
\begin{split}
    \mathcal{L}(\boldsymbol{\theta}) = \frac{1}{N}\frac{1}{\mathcal{C}}\sum_{i=1}^{N}\sum_{k=0}^{\mathcal{C}}&\left[y_k^{(i)}\log(1+e^{-z_k^{(i)}})\right. \\ &\left.+ (1-y_k^{(i)})\log(1+e^{z_k^{(i)}})\right],
    \label{eq:bce}
\end{split}
\end{align}
and training the model $f(\boldsymbol{\theta})$ is to find the optimal weights $\boldsymbol{\theta}_{*}$ by optimizing the loss function in Eq.(\ref{eq:bce}).

\subsection{Distribution-Balanced loss \label{sect:3.3}}

Two practical issues, called ``\textit{label co-occurrence}" and the ``\textit{over-suppression of negative labels}" that make multi-label classification problems more challenging than conventional single-label classification problems. To overcome these challenges, Wu \textit{et al}.~\cite{wu2020distribution} proposed a modified version of the standard BCE loss, namely Distribution-Balanced loss, which consists of two terms: (1) re-balanced weighting and (2) negative-tolerant regularization. The first component, \textit{i.e.}, \textit{re-balance weighting}, was used to tackle the problem of imbalance between classes while taking the co-occurrence of labels into account. Specifically, the \textit{re-balanced weighting} is defined as
\begin{align}
    r_k^{(i)} = \dfrac{P_k^{C}(\textbf{x}^{(i)})}{P^{I}(\textbf{x}^{(i)})},
\end{align}
where $P_k^{C}(\textbf{x}^{(i)})$ and $P^{I}(\textbf{x}^{(i)})$ are the expectation of Class-level sampling frequency and the expectation of Instance-level sampling frequency, respectively. For each image $\textbf{x}^{(i)}$ and class $k$, $n_k = \sum_{i=1}^{N}y_k^{(i)}$ denotes the number of training examples that contain disease class $k$, $P_k^{C}(\textbf{x}^{(i)})$ and $P^{I}(\textbf{x}^{(i)})$ are given as
\begin{align}
    P_k^{C}(\textbf{x}^{(i)}) = \dfrac{1}{\mathcal{C}} \dfrac{1}{n_k},
\end{align}
and 
\begin{align}
    P^{I}(\textbf{x}^{(i)}) = \dfrac{1}{\mathcal{C}} \sum_{y_k^{(i)}=1}^{}\dfrac{1}{n_k}.
\end{align}
To prevent the case where $r$ towards zero and make the training process stable, a smoothing version of the weight
\begin{align}
    \hat{r}_k^{(i)} = \alpha + \frac{1}{1+\exp(-\beta\times(r_k^{(i)}-\mu))}
    \label{eq:r_hat}
\end{align}
is designed to map $r$ into a proper range of values. Here $\alpha$ lifts the value of the weight, while $\beta$ and $\mu$ controls the shape of the mapping function. $\hat{r}_k$ can be adopted to both positive and negative labels although it is initially deduced from positive labels only, in order to preserve class-level consistency. However, we observe that in \cite{wu2020distribution}, the most frequently appearing classes usually have the highest co-existing probability on the condition of other classes. While in the pediatric CXR dataset, the \textit{No Finding} class, the most common class, always presents alone. Thus, in each image $\textbf{x}^{(i)}$ with $y_{\text{No Finding}}^{(i)} = 1$, the re-balancing weight of \textit{No Finding} class is always equal to 1, which is the maximum value of $r$. This will result in not thoroughly eliminate the class imbalance and may even exaggerate it. To address this problem, we propose a modified version of $r_{\text{No Finding}}$ which lowers the impact of \textit{No Finding} samples to the total loss function. Concretely, we define a fixed term
\begin{align}
    \hat{c} = \frac{1}{\mathcal{C}^{2}}\sum_{k=0}^{\mathcal{C}}\dfrac{1}{n_k}.
\end{align}
We then add $\hat{c}$ to the formulation of $r_{\text{No Finding}}$
\begin{align}
    r_{\text{No Finding}}^{(i)} = \dfrac{P_{\text{No Finding}}^{C}(\textbf{x}^{(i)})}{P^{I}(\textbf{x}^{(i)}) + \hat{c}}.
    \label{eq:r_nf}
\end{align}

In multi-label classification problems, an image is usually negative with most classes. Using the standard BCE loss would lead to the over-suppression of the negative side due to its symmetric nature. To tackle this challenge, the second component, namely \textit{negative-tolerant regularization},
\begin{align}
\begin{split}
    \mathcal{L}_{\text{NT}}(\textbf{x}^{(i)},y^{(i)}) &= \frac{1}{\mathcal{C}}\sum_{k=0}^{\mathcal{C}}\left[y_k^{(i)}\log(1+e^{-(z_k^{(i)}-v_k)}) \right. \\ &\left. + \frac{1}{\lambda}(1-y_k^{(i)})\log(1+e^{\lambda(z_k^{(i)}-v_k)})\right]
\end{split}
\end{align}
is constructed, which contains a margin $v$ and a re-scaling factor $\lambda$. Here $v$ is designed by considering intrinsic model bias and played a role of a threshold. The formulation of $v$ is given as
\begin{align}
    v_k = \kappa\log(\frac{N}{n_k}-1),
\end{align}
where $\kappa$ is used as a scale factor to get $v$. We refer the reader to the original work in \cite{wu2020distribution} for more details. The final \textit{Distribution-Balanced loss} is constructed by integrating two components
\begin{align}
\begin{split}
\mathcal{L}_{\text{DB}}(\textbf{x}^{(i)},y^{(i)}) &= \frac{1}{\mathcal{C}}\sum_{k=0}^{\mathcal{C}}\left[y_k^{(i)}\log(1+e^{-(z_k^{(i)}-v_k)}) \right. \\
&\left. \quad +\frac{1}{\lambda}(1-y_k^{(i)})\log(1+e^{\lambda(z_k^{(i)}-v_k)})\right]\hat{r}_k^{(i)},
\end{split}
\end{align}
where $\hat{r}_{\text{No Finding}}$ is calculated by Eq.~(\ref{eq:r_hat}), with $r_{\text{No Finding}}$ is given by Eq.~(\ref{eq:r_nf}).

\subsection{Network architecture \label{sect:3.4}}

Three D-CNNs were exploited for classifying common thoracic diseases in pediatric CXR images, including DenseNet-121~\cite{huang2017densely}, Dense-169~\cite{huang2017densely}, and ResNet-101~\cite{he2016deep}. These networks have achieved significant performance on the ImageNet dataset~\cite{krizhevsky2012imagenet}, a large-scale used to benchmark classification models~\cite{imagenet2014}. More importantly, these network architectures were well-known as the most successful D-CNNs for medical applications, particularly for the CXR interpretation~\cite{rajpurkar2017chexnet,irvin2019chexpert,pham2020interpreting}. For each network, we followed the original implementations~\cite{huang2017densely,he2016deep} with some minor modifications. Specifically, we replaced the final fully connected layer in each network with a fully connected layer producing a 10-dimensional output. We then applied the sigmoid nonlinearity to produce the final output, representing the predicted probability of the presence of each pathology class.
 
\subsection{Training methodology \label{sect:3.5}}

We applied state-of-the-art techniques in training deep neural networks to improve learning performance on the imbalanced pediatric CXR dataset, including transfer learning and ensemble learning. Details are described below

\subsubsection{Transfer learning from adult to pediatric CXR}

Pediatric CXR data is limited due to the high labeling cost and the protocol of limiting children's exposure to radiation. Fortunately, there is a large amount of adult CXR data available that we can leverage. To improve the learning performance on the pediatric CXR, we propose to train D-CNNs on a large-scale adult CXR dataset (source domain) and then finetune the pre-trained networks on our pediatric CXR dataset (target domain). In the experiments, we first trained DenseNet-121~\cite{huang2017densely} on CheXpert~\cite{irvin2019chexpert} -- a large adult CXR dataset that contains 224,316 CXR scans. We then initialized the network with the pre-trained weights and finally finetuned it on the pediatric CXR dataset. An ablation study was conducted to verify the effectiveness of the proposed transfer learning method. Experimental results are reported in Section~\ref{sect:performance}, and Table~\ref{tab:dda_result}. 

\subsubsection{Ensemble learning}

It is hard for a single D-CNN model to obtain a high and consistent performance across all pathology classes in a multi-label classification task. Empirically, the diagnostic accuracy for each pathology often varies and depends on the choice of network architecture. An ensemble learning approach that combines multiple classifiers should be explored to achieve a highly accurate classifier. In this work, we leveraged the power of ensemble learning by combining the predictions of three different pre-trained D-CNNs: DenseNet-121~\cite{huang2017densely}, DenseNet-169~\cite{huang2017densely}, and ResNet-101~\cite{he2016deep}. Concretely, the outputs of the pre-trained networks were concatenated into a prediction vector, and then the averaging operation was used to produce the final prediction.

\subsection{Visual interpretability \label{sect:3.6}}

Explainability is a crucial factor in transferring artificial intelligence (AI) models into clinical practice~\cite{tonekaboni2019clinicians,vellido2019importance}. An interpretable AI system~\cite{samek2019explainable} is able to provide the links between learned features and predictions. Such systems help radiologists understand the underlying reasoning of diagnostic results and identify individual cases for which the predictors potentially give incorrect predictions. In this work, Gradient-weighed Class Activation Mapping (Grad-CAM)~\cite{Selvaraju2019_gradcam} was used to highlight features that strongly correlate with the output of the proposed model. This method aims to stick to the gradient passed through the network to determine the relevant features. Given a convolutional layer $l$ in a trained model, denoting $A_{l}^{k}$ as the activation map for the \textit{k}-th channel, and $Y^c$ as the probability of class $c$. The Grad-CAM $L_{\text{Grad-CAM}}^{c}$, is constructed~\cite{wang2020score} as
\begin{align}
L_{\text{Grad-CAM}}^{c} = ReLU\left( \sum_{k}\alpha_{k}^{c} A_{l}^{k} \right),
\end{align}
where
\begin{align}
\alpha_{k}^{c} = GP \left( \frac{\partial Y_{}^{c}}{\partial A_{l}^{k}} \right),
\end{align}
and $GP(\cdot)$ denotes the global pooling operation.



\section{Experiment and Result}

\subsection{Datasets \& Implementation details \label{sect:dataset}}

\indent\textbf{Data collection} The pediatric CXR dataset used in this study was retrospectively collected from a primary Children's Hospital between the period 2020-2021. The study has been reviewed and approved by the institutional review board (IRB) of the hospital. The need for obtaining informed patient consent was waived because this work did not impact clinical care. The raw data were completely in DICOM format, in which each study contains a single instance. To keep patient's Protected Health Information (PHI) secure, all patient-identifiable information has been removed except several DICOM attributes that are essential for evaluating the lung conditions like patient's age and sex.

\begin{figure*}[ht]
    \begin{center}
   \includegraphics[width=0.9\linewidth]{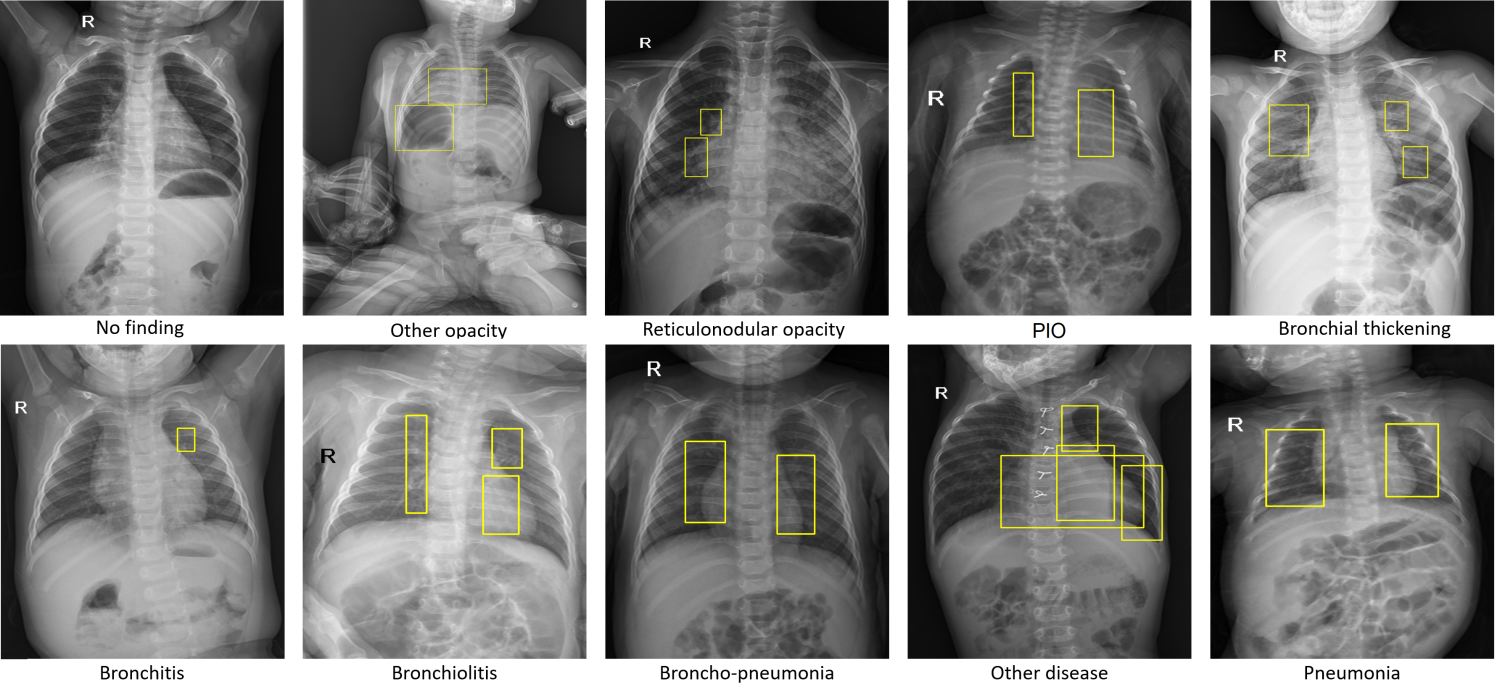}
   \caption{Several representative pediatric CXR images for ``\textit{No finding}'' and  other common lung pathologies in children patients. Bounding box annotations indicate lung abnormalities and are used for visualization purposes. }
   \label{fig:fig1}
   \end{center}
\end{figure*}

\begin{figure}[H]
\begin{center}
   \includegraphics[width=1\linewidth]{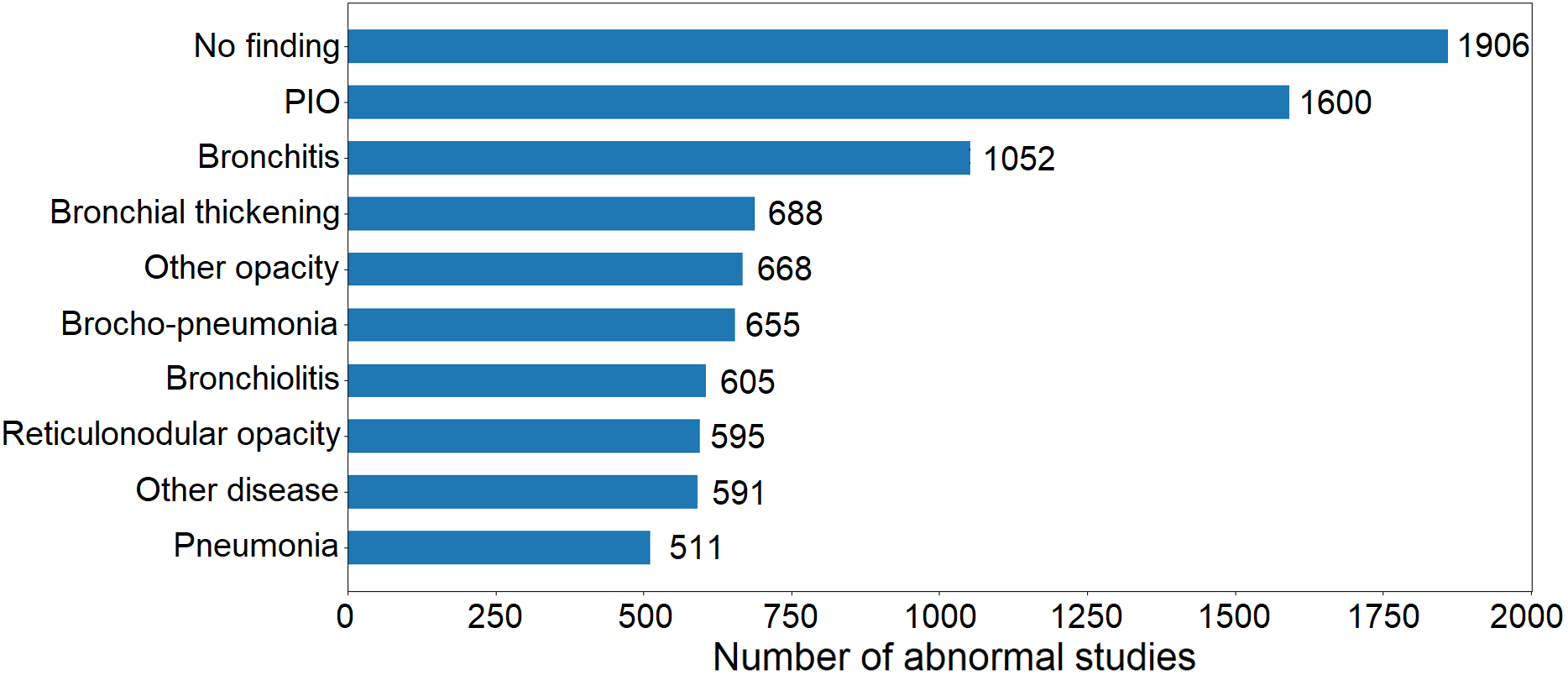}
\end{center}
   \caption{Distribution of disease classes in the whole pediatric CXR dataset used in this study.}\label{fig:class_dist}
\end{figure}

\indent\textbf{Data annotation} A total of 5,017 pediatric CXR scans (normal = 1,906 [37.99\%]; abnormal = 3,111 [62.01\%]) were collected and annotated by a team of expert radiologists who have at least 10 years of experience. During the labeling process, each scan was assigned and notated by one radiologist. The labeling process was performed via an in-house DICOM labeling framework called VinDr Lab (\url{https://vindr.ai/vindr-lab})~\cite{VinDr-Lab}. The dataset was labeled for the presence of 10 pathologies. The ``\textit{No finding}" label was intended to represent the absence of all pathologies. We randomly stratified the dataset into training (70\%), validation (15\%), and test (15\%) sets and ensured that there is no patient overlap between these data sets. The patient characteristics of each data set are summarized in Table~\ref{tab:tab2}. Figure~\ref{fig:fig1} shows several representative pediatric CXR samples from the dataset. The distribution of different disease categories, which reveals the class imbalance problem in the dataset, is shown in Figure~\ref{fig:class_dist}. 

\begin{table*}
\small{
\centering
\label{data-characs}
\begin{tabular}{p{10pt} | p{165pt}|p{60pt}|p{60pt} | p{60pt} | p{60pt}}
\hline
\hline
& \textbf{Variables} & \textbf{Training set} & \textbf{Validation set} & \textbf{Test set} & \textbf{Total}\\
\hline
  \parbox[t]{3mm}{\multirow{5}{*}{\rotatebox[origin=c]{90}{\textbf{Statistics}}}}
& Acquisition time (years)              & 2020 -- 2021      & 2020 -- 2021      & 2020 -- 2021      & 2020 -- 2021 \\
& Age$^{(\dag)}$, mean (range)          & 1.55 (0--10)       & 1.45 (0--10)     & 1.36 (0--10)      &  1.51 (0--10) \\
& Image size, mean                      & 1,639$\times$1,346  & 1,645$\times$1,352  & 1,622$\times$1,339  & 1,637$\times$1,345 \\
& Gender$^{(\dag)}$, male (\%)                     & 60.71             & 60.14             & 57.39             &  60.12 \\
& Number of images                      & 3,550              & 744               & 777               & 5,071 \\
\hline
\parbox[t]{3mm}{\multirow{10}{*}{\rotatebox[origin=c]{90}{\textbf{Pathology}}}}  
& 1. Reticulonodular opacity (\%)                  & 402 (11.32)       & 90 (12.10)        & 103 (13.26)   & 595 (11.73)\\
& 2. Peribronchovascular interstitial opacity (\%) & 1,116 (31.44)      & 232 (31.18)       & 252 (32.43)   & 1,600 (31.55)\\
& 3. Other opacity (\%)                            & 453 (12.76)       & 97 (13.04)        & 118 (15.19)   & 668 (13.17)\\
& 4. Bronchial thickening (\%)                     & 477 (13.44)       & 101 (13.58)       & 110 (14.16)   & 688 (13.57)\\
& 5. Bronchitis (\%)                               & 730 (20.56)       & 161 (21.64)       & 161 (20.72)   & 1,052 (20.75)\\
& 6. Brocho-pneumonia  (\%)                        & 438 (12.34)       & 97 (13.04)        & 120 (15.44)   & 655 (12.92)\\
& 7. Bronchiolitis (\%)                            & 417 (11.75)       & 87 (11.69)        & 101 (13.0)    & 605 (11.93)\\
& 8. Pneumonia (\%)                            & 354 (9.97)        & 72 (9.68)         & 85 (10.94)    & 511 (10.08)\\
& 9. Other disease  (\%)                           & 396 (11.15)       & 87 (11.69)        & 108 (13.9)    & 591 (11.65)\\
& 10. No finding (\%)                                  & 1,387 (39.07)      & 287 (38.58)       & 232 (29.86)   & 1,906 (37.59)\\
\hline
\hline
\end{tabular}
\caption{Demographic data of training, validation, and test sets. ($\dag$) These calculations were performed on the number of studies where gender and age were available.}\label{tab:tab2}
}
\end{table*}

\textbf{Implementation details} To evaluate the effectiveness of the proposed method, several experiments have been conducted. First, we investigated the impact of transfer learning by comparing the model performance when finetuning with pre-trained weights from CheXpert~\cite{irvin2019chexpert}, ImageNet~\cite{imagenet2014}, and training from scratch with random initial weights. We then verified the impact of the ensembling method on the classification performance of the whole framework. For all experiments, we enhanced the contrast of the image by equalizing histogram and then rescaled them to 512$\times$512 resolution before inputting the images into the networks. Model's parameters were updated using stochastic gradient descent (SGD) with a momentum of 0.9. Each network was trained end-to-end for 80 epochs with a total batch size of 32 images. The learning rate was initially set at $1\times10^{-3}$  and updated by the triangular learning rate policy~\cite{smith2017cyclical}. All networks were implemented and trained using Python (v3.7.0) and Pytorch framework (v1.7.1). The hardware we used for the experiments was two NVIDIA RTX 2080Ti 11GB RAM intergrated with the CPU Intel Core i9-9900k 32GB RAM.

\subsection{Evaluation metrics}

The performance of the proposed method was measured using the area under the receiver operating characteristic curve (AUC). The AUC score represents a degree of measure of separability and the higher the AUROC achieves. We also reported sensitivity, specificity and, \textit{F1}-score at the optimal cut-off point. Specifically, the optimal threshold $c^{*}$ of the classifier is determined by maximizing Youden’s index~\cite{youden1950index} $J(c)$ where $J(c) = q(c) + r(c) - 1$. Here the sensitivity $q$ and the specificity $r$ are functions of the cut-off value $c$. To assess the statistical significance of performance indicators, we estimate the 95\% confidence interval (CI) by bootstrapping with 10,000 replications. 

\subsection{Comparison to state-of-the-art}

To demonstrate the effectiveness of the proposed approach, we compared our result with recent state-of-the-art methods for the pediatric CXR interpretation~\cite{Singh2020,Rahman_2020,chen2019deep}. To this end, we reproduced these approaches on our pediatric CXR dataset and reported their performance on the test set (\textit{N = 777}) using the AUC score. For a fair comparison, we applied the same training methodologies and hyper-parameter settings as reported in the original papers~\cite{Singh2020,Rahman_2020,chen2019deep}. We report the experimental results in Section~\ref{sect:performance}, and Table~\ref{tab:compare_sota}.

\subsection{Experimental results \& quantitative analysis}

\subsubsection{Model performance \label{sect:performance}} The mean AUC score of 10 classes of DenseNet-121~\cite{huang2017densely} with different initial weight values is shown in Table~\ref{tab:dda_result}. The model finetuning with pre-trained weights on CheXpert~\cite{irvin2019chexpert} showed the best performance with an AUC of 0.715 (95\% CI, 0.693--0.737), 0.696 (95\% CI, 0.675--0.716) on the validation and test set, respectively. Meanwhile, DenseNet-121~\cite{huang2017densely} trained with random initial weight values reported an AUC of 0.686 (95\% CI, 0.664--0.708) on the validation set, and 0.657 (95\% CI, 0.636--0.678) on the test set, which is the worst performance compared to the other two approaches.

\begin{table}[ht]
\fontsize{10pt}{12pt}
\selectfont
\caption{Mean AUC with different initial weight values for DensetNet-121 on the validation and test sets. Best results are in \textbf{bold}.}
\resizebox{\textwidth}{!}{%
\begin{tabular}{l|c|c}
\hline
\hline
\textbf{Initialization} & \textbf{Validation set} & \textbf{Test set} \\ \hline
Random                                          & 0.673 (0.650-0.696) & 0.657 (0.636-0.679) \\ 
ImageNet                                        & 0.686 (0.664-0.708) & 0.657 (0.636-0.678) \\ 
\hline
\textbf{CheXpert (ours)}                                      & \textbf{0.715 (0.693-0.737)} & \textbf{0.696 (0.675-0.716)} \\ \hline
\hline
\end{tabular}
}
\label{tab:dda_result}
\end{table}

\begin{figure*} [ht]
    \centering 
    \includegraphics[width=1\textwidth]{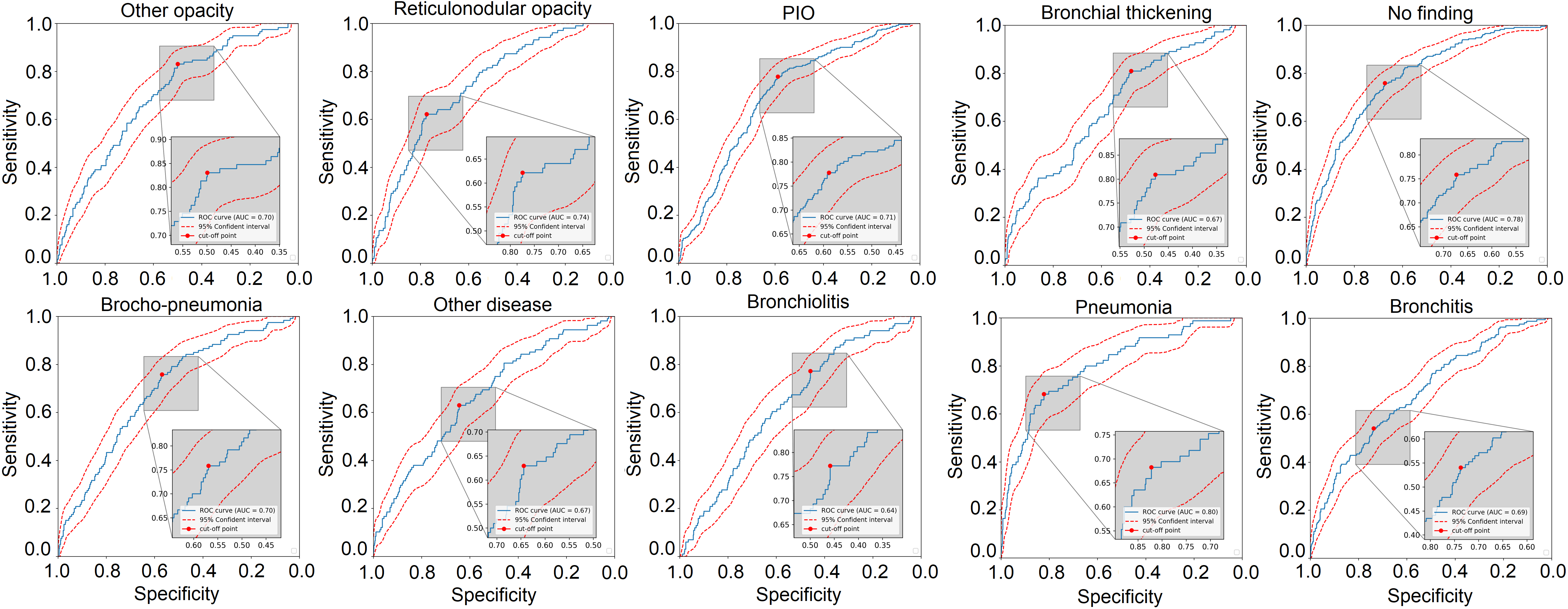}
    \caption{ROC curves of the ensemble model for 10 pathologies on the test set. Best viewed in a computer by zooming-in. }
    \label{fig:auc_curve}
\end{figure*}

Table \ref{tab:ensemble_result} provides a comparison of the classification performance between 3 single models (\textit{i.e.}, DenseNet-121~\cite{huang2017densely}, DenseNet-169~\cite{huang2017densely}, ResNet-101~\cite{he2016deep}) and the ensemble model that combines results of all models. On both the validation and test sets, the ensemble model outperformed all three single models with an AUC of 0.733 (95\% CI, 0.713--0.754) and 0.709 (95\% CI, 0.690--0.729), respectively. The ensemble model's performance for each disease class in the test set is shown in Table \ref{tab:experimental_result}. At the optimal cut-off point, it achieved a sensitivity of 0.722, a specificity of 0.579, and an \textit{F1}-score of 0.389 on the test set. We observed that the reported performances varied over the target diseases, \textit{e.g.}, the final ensemble model performed best on 2 classes \textit{Pneumonia} and \textit{No finding}, while the worst was on \textit{Bronchiolitis} class. The ROC of each disease class is further shown in Figure~\ref{fig:auc_curve}.

\begin{table}[H]
\fontsize{10pt}{12pt}
\selectfont
\caption{Mean AUC score of single architectures and the ensemble model on the validation and test sets.}
\resizebox{\textwidth}{!}{%
\begin{tabular}{l|c|c}
\hline
\hline
\textbf{Model} & \textbf{Validation set}                                     & \textbf{Test  set}                                   \\ \hline
DenseNet-121~\cite{huang2017densely}                                                          & 0.715 (0.693-0.737)                      & 0.696 (0.675-0.716)                      \\ 
DenseNet-169~\cite{huang2017densely}                                                         & 0.721 (0.699-0.741)                      & 0.691 (0.672-0.711)                      \\ 
ResNet-101~\cite{he2016deep}                                                            & 0.717 (0.696-0.737)                      & 0.700 (0.680-0.719)                      \\ 
\hline
\textbf{Ensemble} & \textbf{0.733 (0.713-0.754)} & \textbf{0.709 (0.690-0.729)} \\ \hline
\hline
\end{tabular}
}
\label{tab:ensemble_result}
\end{table}

\begin{table*}
\fontsize{10pt}{12pt}
\selectfont
\centering
\caption{Experimental results on the validation dataset and comparison with the state-of-the-art. The proposed method outperforms other previous methods on most pathologies in our dataset. Here we highlight the best result in \textcolor{red}{\textbf{red}} and the second-best in \textcolor{blue}{\textbf{blue}}.}
\label{tab:compare_sota}
\resizebox{0.9\textwidth}{!}{%
\begin{tabular}{lcccc}
\hline
\hline
\textbf{Pathology} & \textbf{Chouhan \textit{et al.} \cite{Singh2020} (2020)} & \textbf{Rahman \textit{et al.} \cite{Rahman_2020} (2020)} & \textbf{Chen \textit{et al.} \cite{chen2019deep} (2020)} & \textbf{Proposed method} \\ \hline 
Other opacity & \textcolor{blue}{\textbf{0.6737}} & 0.636 & 0.656 & \textcolor{red}{\textbf{0.703}} \\
Reticulonodular opacity & 0.6870 & 0.652 & \textcolor{blue}{\textbf{0.701}} & \textcolor{red}{\textbf{0.739}}\\
PIO & \textcolor{blue}{\textbf{0.6624}} & 0.619 & 0.653 & \textcolor{red}{\textbf{0.706}} \\
Bronchial   thickening  & \textcolor{blue}{\textbf{0.6791}} & 0.648 & 0.647 & \textcolor{red}{\textbf{0.673}}\\
No   finding & \textcolor{blue}{\textbf{0.7740}} & 0.734 & 0.746 & \textcolor{red}{\textbf{0.776}} \\
Bronchitis & \textcolor{blue}{\textbf{0.6613}} & 0.648 & 0.652 & \textcolor{red}{\textbf{0.691}}\\
Brocho-pneumonia & 0.6710 & 0.648 & \textcolor{blue}{\textbf{0.677}} & \textcolor{red}{\textbf{0.696}}\\
Other   disease & \textcolor{blue}{\textbf{0.6754}} & 0.581 & 0.653 & \textcolor{red}{\textbf{0.669}}\\
Bronchiolitis & 0.6089 & 0.639 & \textcolor{red}{\textbf{0.648}} & \textcolor{blue}{\textbf{0.638}}\\
Pneumonia & 0.7097 & 0.682 & \textcolor{blue}{\textbf{0.737}} & \textcolor{red}{\textbf{0.802}}\\ \hline
Mean & \textcolor{blue}{\textbf{0.6802}} & 0.649 & 0.677 & \textcolor{red}{\textbf{0.709}}\\ \hline
\hline
\end{tabular}
}
\label{tab:comparison}
\end{table*}

\begin{table}[H]
\fontsize{10pt}{12pt}
\centering
\caption{Performance of the ensemble model for each disease class on the test set.}
\resizebox{0.95\textwidth}{!}{%
\begin{tabular}{l|c|c|c|c}
\hline
\hline
\textbf{Label} & \textbf{AUROC}                & \textbf{Sensitivity}       & \textbf{Specificity}         & \textbf{\textit{F1}-score}                  \\ \hline
Other   opacity                                                       & 0.703              & 0.856             & 0.373               & 0.320                \\
Reticulonodular   opacity                                             & 0.739              & 0.786             & 0.559               & 0.337               \\
PIO$^{(*)}$                         & 0.706              & 0.817             & 0.522               & 0.581               \\
Bronchial   thickening                                                & 0.673              & 0.627             & 0.571               & 0.297               \\
No   finding                                                          & 0.776              & 0.668             & 0.739               & 0.586               \\
Bronchitis                                                            & 0.691              & 0.571             & 0.69                & 0.414               \\
Brocho-pneumonia                                                      & 0.696              & 0.725             & 0.581               & 0.361               \\
Other   disease                                                       & 0.669              & 0.806             & 0.445               & 0.307               \\
Bronchiolitis                                                         & 0.638              & 0.683             & 0.496               & 0.270                \\
Pneumonia                                                             & 0.802              & 0.682             & 0.809               & 0.422               \\ \hline
\textbf{Mean}                                                         & \textbf{0.709} & \textbf{0.722} & \textbf{0.579} & \textbf{0.389} \\\hline \hline
\end{tabular}
}
\label{tab:experimental_result}
\end{table}


\renewcommand\figurename{Figure}

\subsubsection{Effect of modified Distribution-Balanced loss} 

We conducted ablation studies on the effect of the modified Distribution-Balanced loss. Specifically, we reported the diagnostic accuracy of DenseNet-121~\cite{huang2017densely} on our pediatric CXR test set when trained with the modified Distribution-Balanced loss and other standard losses, including the BCE loss, weighted BCE loss~\cite{rajpurkar2017chexnet}, Focal loss~\cite{lin2018focal}, and the original Distribution-Balanced (DB) loss~\cite{wu2020distribution}. For all experiments, we used the same hyperparameter setting for network training. Table~\ref{tab:ablation_study} shows the result of this experiment. The network trained with the modified Distribution-Balanced loss achieved an AUC of 0.683 (95\% CI, 0.662--0.703) and a \textit{F1}-score of 0.368 (95\% CI, 0.350--0.385), respectively. These results outperformed all other standard losses with  large margins. For instance, our approach showed an improvement of 1.3\% in AUC and of 0.4\% in \textit{F1}-score compared to the second-best results. These improvements validated the effectiveness of the modified Distribution-Balanced loss in learning disease patterns from the unbalanced pediatric CXR dataset.

\begin{table}[H]
\fontsize{10pt}{12pt}
\selectfont
\centering
\caption{Performance of the DenseNet-121~\cite{huang2017densely} on the test set of our pediatric CXR dataset using different loss functions.}
\resizebox{0.95\textwidth}{!}{%
\begin{tabular}{l|c|c}
\hline
\hline
\textbf{Loss} & \textbf{AUROC} & \textbf{\textit{F1}-score}      \\ \hline
BCE         & 0.657 (0.636-0.678)   & 0.346 (0.328-0.364) \\
Weighted-BCE~\cite{rajpurkar2017chexnet}       & \textcolor{blue}{\textbf{0.670 (0.650-0.691)}} & 0.354 (0.336-0.371) \\
Focal loss~\cite{lin2018focal}  & 0.668 (0.647-0.689)   & 0.355 (0.338-0.371)\\
DB loss~\cite{wu2020distribution}     & 0.665 (0.644-0.686)   & \textcolor{blue}{\textbf{0.363 (0.345-0.380)}}
\\
\hline
\textbf{Ours} & \textcolor{red}{\textbf{0.683 (0.662-0.703)}} & \textcolor{red}{\textbf{0.368 (0.350-0.385)}}\\
\hline
\hline
\end{tabular}
}
\label{tab:ablation_study}
\end{table}

\subsubsection{Model interpretation}

\begin{figure}[H]
\begin{center}
   \includegraphics[width=1\linewidth]{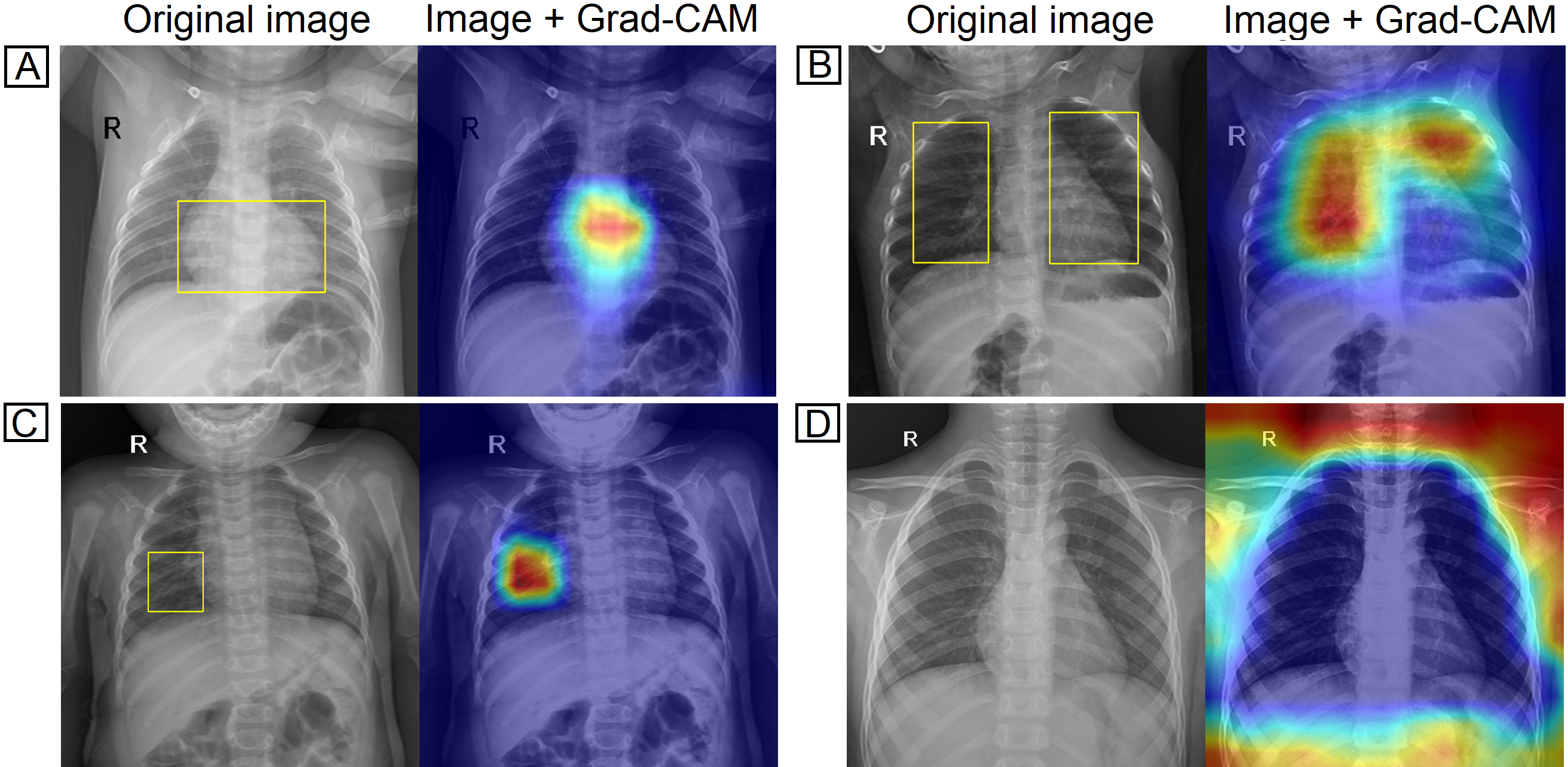}
\end{center}
   \caption{Saliency maps indicated the regions of each radiograph with the most significant influence on the models’ prediction.}\label{fig:gradcam}
\end{figure}

We computed Grad-CAM~\cite{Selvaraju2019_gradcam} to visualize the areas of the radiograph which the network predicted to be most indicative of each disease. Saliency maps generated  by Grad-CAM were then rescaled to match the dimensions of the original images and overlay the map on the images. Figure~\ref{fig:gradcam}(A--C) shows some pediatric CXR scans with different respiratory pathologies, while Figure~\ref{fig:gradcam}D represents a normal lung. Heatmap images are provided alongside the ground-truth boxes annotated by board-certified radiologists. As we can see, the trained models can localize the regions that have lesions in positive cases and shows no focus on the lung region in negative cases.

\section{Conclusion}

In this paper, we introduced a deep learning-based approach to detect common pulmonary pathologies on CXR of pediatric patients. To the best of our knowledge, this is the first effort to address the classification of multiple diseases from pediatric CXRs. In particular, we proposed modifying the Distribution-Balanced loss to reduce the impact of class imbalance in classification performance. Our experiments demonstrated the effectiveness of the proposed method. Although the proposed system surpassed previous state-of-the-art approaches, we recognized that its performance remains low compared to the human expert performance. This reveals the major challenge in learning disease features on pediatric CXR images using deep learning techniques, opening new aspects for future research. Future works include developing a localization model for identifying abnormalities on the pediatric CXR scans and investigating the impact of the proposed deep learning system on clinical practice.

{\small
\bibliographystyle{ieee}
\bibliography{egbib}

\begin{thebibliography}{10}\itemsep=-1pt

\bibitem{baltruschat2019comparison}
I.~M. Baltruschat, H.~Nickisch, M.~Grass, T.~Knopp, and A.~Saalbach.
\newblock {Comparison of deep learning approaches for multi-label chest X-ray
  classification}.
\newblock {\em Scientific Reports}, 9(1):1--10, 2019.

\bibitem{chen2019deep}
H.~Chen, S.~Miao, D.~Xu, G.~D. Hager, and A.~P. Harrison.
\newblock {Deep hierarchical multi-label classification of chest X-ray images}.
\newblock In {\em International Conference on Medical Imaging with Deep
  Learning}, pages 109--120, 2019.

\bibitem{Chen2020}
K.-C. Chen, H.-R. Yu, W.-S. Chen, W.-C. Lin, Y.-C. Lee, H.-H. Chen, J.-H.
  Jiang, T.-Y. Su, C.-K. Tsai, T.-A. Tsai, C.-M. Tsai, and H.~Lu.
\newblock {Diagnosis of common pulmonary diseases in children by X-ray images
  and deep learning}.
\newblock {\em Scientific Reports}, 10(1):1--9, 2020.

\bibitem{GBD2015}
G.~.~L. Collaborators.
\newblock Estimates of the global, regional, and national morbidity, mortality,
  and aetiologies of lower respiratory tract infections in 195 countries: a
  systematic analysis for the global burden of disease study 2015.
\newblock {\em The Lancet Infectious Diseases}, 17(11):1133--1161, 2017.

\bibitem{imagenet2014}
J.~Deng, W.~Dong, R.~Socher, L.-J. Li, K.~Li, and L.~Fei-Fei.
\newblock {ImageNet: A large-scale hierarchical image database}.
\newblock In {\em IEEE Conference on Computer Vision and Pattern Recognition},
  pages 248--255, 2009.

\bibitem{du2002observer}
G.~Du~Toit, G.~Swingler, and K.~Iloni.
\newblock Observer variation in detecting lymphadenopathy on chest radiography.
\newblock {\em International Journal of Tuberculosis and Lung Disease},
  6(9):814--817, 2002.

\bibitem{ganda2018survey}
D.~Ganda and R.~Buch.
\newblock A survey on multi label classification.
\newblock {\em Recent Trends in Programming Languages}, 5(1):19--23, 2018.

\bibitem{gu2018classification}
X.~Gu, L.~Pan, H.~Liang, and R.~Yang.
\newblock Classification of bacterial and viral childhood pneumonia using deep
  learning in chest radiography.
\newblock In {\em Proceedings of the International Conference on Multimedia and
  Image Processing}, pages 88--93, 2018.

\bibitem{he2016deep}
K.~He, X.~Zhang, S.~Ren, and J.~Sun.
\newblock Deep residual learning for image recognition.
\newblock In {\em IEEE Conference on Computer Vision and Pattern Recognition},
  pages 770--778, 2016.

\bibitem{huang2017densely}
G.~Huang, Z.~Liu, L.~Van Der~Maaten, and K.~Q. Weinberger.
\newblock Densely connected convolutional networks.
\newblock In {\em IEEE Conference on Computer Vision and Pattern Recognition},
  pages 4700--4708, 2017.

\bibitem{ibrahim2020confidence}
K.~M. Ibrahim, E.~V. Epure, G.~Peeters, and G.~Richard.
\newblock Confidence-based weighted loss for multi-label classification with
  missing labels.
\newblock In {\em Proceedings of the International Conference on Multimedia
  Retrieval}, pages 291--295, 2020.

\bibitem{irvin2019chexpert}
J.~Irvin, P.~Rajpurkar, M.~Ko, Y.~Yu, S.~Ciurea-Ilcus, C.~Chute, H.~Marklund,
  B.~Haghgoo, R.~Ball, K.~Shpanskaya, et~al.
\newblock {CheXpert: A large chest radiograph dataset with uncertainty labels
  and expert comparison}.
\newblock In {\em AAAI Conference on Artificial Intelligence}, volume~33, pages
  590--597, 2019.

\bibitem{krizhevsky2012imagenet}
A.~Krizhevsky, I.~Sutskever, and G.~E. Hinton.
\newblock Imagenet classification with deep convolutional neural networks.
\newblock {\em Advances in Neural Information Processing Systems},
  25:1097--1105, 2012.

\bibitem{Labhane2020}
G.~{Labhane}, R.~{Pansare}, S.~{Maheshwari}, R.~{Tiwari}, and A.~{Shukla}.
\newblock {Detection of pediatric pneumonia from chest X-ray images using CNN
  and transfer learning}.
\newblock In {\em International Conference on Emerging Technologies in Computer
  Engineering: Machine Learning and Internet of Things}, pages 85--92, 2020.

\bibitem{liang2020transfer}
G.~Liang and L.~Zheng.
\newblock A transfer learning method with deep residual network for pediatric
  pneumonia diagnosis.
\newblock {\em Computer Methods and Programs in Biomedicine}, 187:104964, 2020.

\bibitem{lin2018focal}
T.-Y. Lin, P.~Goyal, R.~Girshick, K.~He, and P.~Doll{\'a}r.
\newblock Focal loss for dense object detection.
\newblock In {\em Proceedings of the IEEE International Conference on Computer
  Vision}, pages 2980--2988, 2017.

\bibitem{liu2020emerging}
W.~Liu, X.~Shen, H.~Wang, and I.~W. Tsang.
\newblock The emerging trends of multi-label learning.
\newblock {\em arXiv preprint arXiv:2011.11197}, 2020.

\bibitem{moore2019machine}
M.~M. Moore, E.~Slonimsky, A.~D. Long, R.~W. Sze, and R.~S. Iyer.
\newblock Machine learning concepts, concerns and opportunities for a pediatric
  radiologist.
\newblock {\em Pediatric Radiology}, 49(4):509--516, 2019.

\bibitem{VinDr-Lab}
N.~T. Nguyen, P.~T. Truong, V.~T. Ho, T.~V. Nguyen, H.~T. Pham, M.~T. Nguyen,
  L.~T. Dam, and H.~Q. Nguyen.
\newblock {VinDr Lab: A Data Platform for Medical AI}.
\newblock \url{https://github.com/vinbigdata-medical/vindr-lab}, 2021.

\bibitem{pham2020interpreting}
H.~H. Pham, T.~T. Le, D.~Q. Tran, D.~T. Ngo, and H.~Q. Nguyen.
\newblock {Interpreting chest X-rays via CNNs that exploit hierarchical disease
  dependencies and uncertainty labels}.
\newblock {\em Neurocomputing}, 437:186--194, 2021.

\bibitem{Rahman_2020}
T.~Rahman, M.~E.~H. Chowdhury, A.~Khandakar, K.~R. Islam, K.~F. Islam, Z.~B.
  Mahbub, M.~A. Kadir, and S.~Kashem.
\newblock {Transfer learning with deep convolutional neural network (CNN) for
  pneumonia detection using chest X-ray}.
\newblock {\em Applied Sciences}, 10(9):3233, May 2020.

\bibitem{rajaraman2019visualizing}
S.~Rajaraman, S.~Candemir, G.~Thoma, and S.~Antani.
\newblock Visualizing and explaining deep learning predictions for pneumonia
  detection in pediatric chest radiographs.
\newblock In {\em Medical Imaging 2019: Computer-Aided Diagnosis}, volume
  10950, page 109500S. International Society for Optics and Photonics, 2019.

\bibitem{Rajaraman2019}
S.~Rajaraman, S.~Candemir, G.~Thoma, and S.~Antani.
\newblock {Visualizing and explaining deep learning predictions for pneumonia
  detection in pediatric chest radiographs}.
\newblock In K.~Mori and H.~K. Hahn, editors, {\em Medical Imaging 2019:
  Computer-Aided Diagnosis}, volume 10950, pages 200 -- 211, 2019.

\bibitem{rajpurkar2018deep}
P.~Rajpurkar, J.~Irvin, R.~L. Ball, K.~Zhu, B.~Yang, H.~Mehta, T.~Duan,
  D.~Ding, A.~Bagul, C.~P. Langlotz, et~al.
\newblock {Deep learning for chest radiograph diagnosis: A retrospective
  comparison of the CheXNeXt algorithm to practicing radiologists}.
\newblock {\em PLoS Medicine}, 15(11):e1002686, 2018.

\bibitem{rajpurkar2017chexnet}
P.~Rajpurkar, J.~Irvin, K.~Zhu, B.~Yang, H.~Mehta, T.~Duan, D.~Ding, A.~Bagul,
  C.~Langlotz, K.~Shpanskaya, et~al.
\newblock {CheXNet: Radiologist-level pneumonia detection on chest x-rays with
  deep learning}.
\newblock {\em arXiv preprint arXiv:1711.05225}, 2017.

\bibitem{samek2019explainable}
W.~Samek, G.~Montavon, A.~Vedaldi, L.~K. Hansen, and K.-R. M{\"u}ller.
\newblock {\em {Explainable AI: Interpreting, explaining and visualizing deep
  learning}}, volume 11700.
\newblock Springer Nature, 2019.

\bibitem{Selvaraju2019_gradcam}
R.~R. Selvaraju, M.~Cogswell, A.~Das, R.~Vedantam, D.~Parikh, and D.~Batra.
\newblock {Grad-CAM: Visual explanations from deep networks via gradient-based
  localization}.
\newblock {\em International Journal of Computer Vision}, 128(2):336–359, Oct
  2019.

\bibitem{Sharma2020}
H.~{Sharma}, J.~S. {Jain}, P.~{Bansal}, and S.~{Gupta}.
\newblock Feature extraction and classification of chest x-ray images using cnn
  to detect pneumonia.
\newblock In {\em International Conference on Cloud Computing, Data Science
  Engineering}, pages 227--231, 2020.

\bibitem{Singh2020}
S.~Singh, A.~Khamparia, D.~Gupta, P.~Tiwari, C.~Moreira, R.~Damasevicius, and
  V.~Albuquerque.
\newblock {A novel transfer learning based approach for pneumonia detection in
  chest X-ray images}.
\newblock {\em Applied Sciences}, 10:559, 01 2020.

\bibitem{smith2017cyclical}
L.~N. Smith.
\newblock Cyclical learning rates for training neural networks.
\newblock In {\em IEEE Winter Conference on Applications of Computer Vision},
  pages 464--472. IEEE, 2017.

\bibitem{swingler2005diagnostic}
G.~Swingler, G.~Du~Toit, S.~Andronikou, L.~Van~der Merwe, and H.~Zar.
\newblock Diagnostic accuracy of chest radiography in detecting mediastinal
  lymphadenopathy in suspected pulmonary tuberculosis.
\newblock {\em Archives of Disease in Childhood}, 90(11):1153--1156, 2005.

\bibitem{taylor2018automated}
A.~G. Taylor, C.~Mielke, and J.~Mongan.
\newblock {Automated detection of moderate and large pneumothorax on frontal
  chest X-rays using deep convolutional neural networks: A retrospective
  study}.
\newblock {\em PLoS Medicine}, 15(11):e1002697, 2018.

\bibitem{tonekaboni2019clinicians}
S.~Tonekaboni, S.~Joshi, M.~D. McCradden, and A.~Goldenberg.
\newblock What clinicians want: contextualizing explainable machine learning
  for clinical end use.
\newblock In {\em Machine Learning for Healthcare Conference}, pages 359--380,
  2019.

\bibitem{tsoumakas2007multi}
G.~Tsoumakas and I.~Katakis.
\newblock {Multi-label classification: An overview}.
\newblock {\em International Journal of Data Warehousing and Mining},
  3(3):1--13, 2007.

\bibitem{vellido2019importance}
A.~Vellido.
\newblock The importance of interpretability and visualization in machine
  learning for applications in medicine and health care.
\newblock {\em Neural Computing and Applications}, pages 1--15, 2019.

\bibitem{wang2020score}
H.~Wang, Z.~Wang, M.~Du, F.~Yang, Z.~Zhang, S.~Ding, P.~Mardziel, and X.~Hu.
\newblock {Score-CAM: Score-weighted visual explanations for convolutional
  neural networks}.
\newblock In {\em Proceedings of the IEEE Conference on Computer Vision and
  Pattern Recognition Workshops}, pages 24--25, 2020.

\bibitem{unicef2006}
T.~M. Wardlaw, E.~W. Johansson, M.~Hodge, W.~H. Organization, and U.~N. C.~F.
  (UNICEF).
\newblock Pneumonia : the forgotten killer of children, 2006.

\bibitem{wu2020distribution}
T.~Wu, Q.~Huang, Z.~Liu, Y.~Wang, and D.~Lin.
\newblock Distribution-balanced loss for multi-label classification in
  long-tailed datasets.
\newblock In {\em European Conference on Computer Vision}, pages 162--178,
  2020.

\bibitem{youden1950index}
W.~J. Youden.
\newblock Index for rating diagnostic tests.
\newblock {\em Cancer}, 3(1):32--35, 1950.

\bibitem{zar2014global}
H.~Zar and T.~Ferkol.
\newblock The global burden of respiratory disease-impact on child health.
\newblock {\em Pediatric Pulmonology}, 49, 05 2014.

\bibitem{zhang2013review}
M.-L. Zhang and Z.-H. Zhou.
\newblock A review on multi-label learning algorithms.
\newblock {\em IEEE Transactions on Knowledge and Data Engineering},
  26(8):1819--1837, 2013.

\end{thebibliography}
}

\end{document}